\documentclass[11pt]{article}
\usepackage{epsfig}
\usepackage{graphicx,graphics}
\pdfoutput=1
\begin{document}
\pagestyle{myheadings}
%\markright{Neutrino Emission }
\thispagestyle{empty}
\normalsize
\title{{\bf \sc Supermassive Black Holes in the Early Universe}\\
\vspace{0.5cm}
J.A. de Freitas Pacheco\\
\vspace{0.5cm}
Universit\'e de la C\^ote d'Azur\\
Observatoire de la C\^ote d'Azur - Laboratoire Lagrange\\
06304 Nice Cedex - France\\
\vspace{0.5cm}
{\small email:pacheco@oca.eu\\
Keywords: Black holes, quasars, accretion disks}}
%palavras chaves: Expans\~ao C\'osmica, Cosmologia, Modelos Cosmol\'ogicos, Gravita\c{c}\~ao\\
%Section - Physical Sciences}

\date {\today}

\maketitle 
      
%\newpage

\begin{abstract}

The discovery of high redshift quasars represents a challenge to the origin of supermassive black holes. Here, two
evolutionary scenarios are considered. The first one concerns massive black holes in the local universe, which in a
large majority have been formed by the growth of seeds as their host galaxies are assembled in accordance with
the hierarchical picture. In the second scenario, seeds with masses around 100-150~$M_\odot$ grow by accretion of
gas forming a non-steady massive disk, whose existence is supported by the detection of huge amounts of gas and dust 
in high-z quasars. These models of non-steady self-gravitating disks explain quite well the observed ''Luminosity-Mass"
relation of quasars at high-z, indicating also that these objects do not radiate at the so-called Eddington limit.

\end{abstract}

\newpage

\section{Introduction}

The cosmological nature of quasars (QSOs) was established in the early sixties (Schmidt 1963). An immediate consequence of
the implied large distances for these objects was the realization that QSOs were among the most powerful energy sources in the 
universe. Their luminosities are typically around $10^{46} erg.s^{-1}$ but the emission of some QSO's may exceed by one or two orders of magnitude 
that value. Edwin Salpeter was one of the first to propose that a supermassive black hole (SMBH) in a state of accretion could provide the 
necessary energy to explain the luminosities of QSOs (Salpeter 1964). If presently a large majority of the scientific community accepts that
accreting SMBHs are the engines powering QSOs, a series of questions still remain to be answered. For instance, how these SMBHs
are formed? If they grow by accretion what are the seeds and from where they come from? What is the gas accretion geometry: spherically or that of 
an inspiraling disk? In last case, what are the viscous mechanisms responsible for the transfer of angular momentum? 

In the local universe the presence of SMBHs in the center either of elliptical galaxies or bulges of spiral galaxies seems to be a 
well-established fact (Kormendy \& Richstone 1995; Richstone et al. 1998; Kormendy \& Gebhardt 2001). The black hole mass $M$ is well 
correlated either with the stellar mass or the luminosity of the host bulge (Kormendy \& Richstone 1995; Magorrian et al. 1998; 
Marconi \& Hunt 2003; Haring \& Rix 2004; Graham 2007) but, in particular, a tight correlation exists between the SMBH mass and 
the central projected stellar velocity dispersion $\sigma$ (Ferrarese \& Merrit 2000; Gebhardt et al. 2000; Merrit \& Ferrarese 2001; Tremaine et al. 2002). The mechanism 
(or mechanisms) responsible for establishing the M-$\sigma$ relation is (are) not yet well determined but several scenarios have 
been put forward in the past years to explain the origin of such a relation. Self-regulated growth of black holes by feedback effects produced 
either by outflows or UV-radiation from QSO's, which affect also the star formation activity is a possible mechanism able 
to reproduce the M-$\sigma$ relation (Silk \& Rees 1998; Sazonov et al. 2005). A relation between these physical quantities can 
also be obtained from the picture developed by Burkert \& Silk (2001) in which black holes grow at the expense of a viscous 
accretion disk and whose gas reservoir beyond the BH influence radius feeds also the formation of stars.

These investigations seem to point to a well-defined road leading to the formation of SMBHs: the growth of ''seeds'' by accretion 
inside the host galaxy. This picture is consistent with the fact that the present BH mass density agrees with the accreted 
(baryonic) mass density derived from the bolometric luminosity function of quasars (Soltan 1982; Small \& Blandford 1992; Hopkins, 
Richards \& Hernquist 2007) and with a negligible amount of accreted dark matter (Peirani \& de Freitas Pacheco 2008). Seeds 
could be intermediate mass ($10^3-10^4)~M_\odot$ black holes formed during the collapse of primordial gas clouds 
(Haehnelt \& Rees 1993; Eisenstein \& Loeb 1995; Koushiappas, Bullock \& Dekel 2004) or during the core collapse of relativistic star 
clusters formed in star-bursts, which may have occurred in the early evolution of galaxies (Shapiro 2004). Here, as it will be discussed later,
seeds are assumed to be black holes with masses around 100-500~$M_\odot$ originated from the first generation of stars, supposed to be quite
massive due to the absence of metals, which are the main contributors to the cooling of the gas.

The different correlations between the black hole mass and the dynamic or the photometric properties of the host galaxy suggest a gradual
growth of the seed as the host galaxy itself is assembled. However this scenario seems to be inconsistent with the fact that up to now
more than 40 bright QSO's have been discovered at high redshift (Wu et al. 2015). The three QSOs having the highest redshift are
J1061+3922 at z = 6.61, J1120+0641 at z = 7.08 and J1342+0928 at z = 7.54. The later corresponds to an age of the universe of only 0.69 Gyr. Since
most of these high redshift QSOs are associated with SMBHs having masses around $10^8-10^9 ~ M_\odot$, their growth was probably not gradual 
but rather fast in other they could shine so early in the history of the universe. Thus, a possible issue is to admit the existence of
two evolutionary paths leading to the formation of SMBHs: one in which seeds grow intermittently as their host galaxies are assembled and 
another in which seeds grow very fast
in a timescale of less than 1 Gyr. These two possibilities will be discussed in the next sections of this article.

\section{Spherical Accretion and the Eddington limit}

Many authors still consider in their investigations spherical accretion processes in which the mass inflow rate is controlled by the 
Eddington luminosity. In this case, it seems judicious to recall the physical assumptions that permit the derivation of the Eddington
limit. The Euler equation describing a spherically symmetric inflow under the influence of gravitation and radiation pressure is
\begin{equation}
\label{euler}
V\frac{dV}{dr}+\frac{1}{\rho}\left[\frac{d(P+P_r)}{dr}\right]+\frac{GM}{r^2}=0
\end{equation}
In the above equation $V$ is the radial flow velocity, $P$ and $P_r$ are respectively the gas and the radiation pressure, $\rho$ is
the gas density, $G$ is the
gravitational constant and $M$ is the mass of the central object. The radial gradient due to the radiation pressure is given by
\begin{equation}
\frac{dP_r}{dr}=-\frac{1}{c}\int^{\infty}_{0}\kappa_{\nu}\phi_{\nu}d\nu = -\frac{\kappa}{c}\phi
\end{equation}
where $\kappa$ is a suitable frequency average of the total absorption coefficient (including scattering) and $\phi$ is the total 
radiative flux. If the accreting gas envelope is highly ionized, the absorption of photons is essentially due to the Thomson scattering
and, in this case
\begin{equation}
\kappa = \frac{\sigma_T}{\mu m_H}\rho
\end{equation}
where $\sigma_T = 6.65\times 10^{-25}~cm^2$ is the Thomson cross-section, $\mu$ is the mean molecular weight and $m_H$ is the proton mass. If
the medium is optically thin and the radiation comes essentially from the deep inside region of the envelope, then the radiative flux
in the outer regions is simply
\begin{equation}
\phi = \frac{L}{4\pi r^2}
\end{equation}
Combine eqs. (2), (3), (4) and replace into the Euler equation to obtain
\begin{equation}
V\frac{dV}{dr}+\frac{1}{\rho}\frac{dP}{dr}= -\frac{GM}{r^2}+\frac{\sigma_T L}{4\pi\mu m_H c r^2}
\end{equation}
Define now the Eddington luminosity as
\begin{equation}
\label{eddington}
L_E = \frac{4\pi GM\mu m_H c}{\sigma_T} = 1.76 \times 10^{38}\left(\frac{M}{M_\odot}\right)~erg.s^{-1}
\end{equation} 
In this case, eq. (5) can be rewritten as
\begin{equation}
\label{euler2}
V\frac{dV}{dr}+\frac{1}{\rho}\frac{dP}{dr}= -\frac{GM(1-\Gamma)}{r^2}
\end{equation} 
where $\Gamma = L/L_E$.

Assume that the gas equation of state is given by $P = K\rho^{\gamma}$ and define also the adiabatic sound velocity as
$a^2 = \gamma(P/\rho)$. Under these conditions, using the mass conservation equation to express the mass density gradient, after 
some algebra, eq.\ref{euler2} can be recast as 
\begin{equation}
\label{euler3}
V\left(1-\frac{a^2}{V^2}\right)\frac{dV}{dr} = \frac{2a^2}{r}-\frac{GM(1-\Gamma)}{r^2}
\end{equation}

The ''critical'' point of the flow in the usual mathematical sense corresponds to the point where both sides of eq.~\ref{euler3} vanish. 
Hence, in order to have the continuity of the flow through the critical point, two conditions must be simultaneously satisfied, namely
\begin{equation}
\label{criticalvelocity}
V_* = a_*
\end{equation}
and
\begin{equation}
\label{criticalradius}
r_* = \frac{GM(1-\Gamma)}{2a_*^2}
\end{equation}
for the critical velocity and the critical radius respectively. These relations imply that the critical and the sonic points of the flow coincide (this
is not always the case) and that the luminosity radiated from inside must be {\bf{smaller}} than the Eddington value in order that the critical
radius be real. Note that once the critical point is surpassed, the left side of eq. \ref{euler3} is negative, requiring imperatively
that the right side be also negative or, equivalently, that $\Gamma < 1$. In other words, the spherical accretion of an optically thin
envelope requires sub-Eddington conditions otherwise the inflow cannot be established. As we will see later, this requirement is weakened
when the inflow geometry is modified as, for instance, in the case of an accretion disk.

As it was mentioned previously, many authors assume that the central black hole accretes mass with the envelope radiating near the Eddington
limit. Since the Eddington luminosity is proportional to the mass of the black hole (see eq. 6), it results an exponential growth with a 
timescale
\begin{equation}
\tau_E = \frac{\eta}{(1-\eta)}\frac{c\sigma_T}{4\pi\mu m_H G} = 3.22\times 10^8\frac{\eta}{(1-\eta)}~yr
\end{equation}
where $\eta$ is the accretion efficiency. Such a short timescale in often considered as an argument to explain the presence of SMBHs at
high-z. However, as we have seen above, the Eddington limit is derived under conditions in which the envelope is optically thin and the opacity is due only
to the Thomson scattering. The optical depth of the envelope is given by
\begin{equation}
\tau(r,\infty) = \int^{\infty}_r \frac{\sigma_T}{\mu m_H}\rho(r')dr' = 7.8\times 10^{-6}\left(\frac{M}{M_\odot}\right)n_{\infty}
\end{equation}
and the gravitational radius was taken as a lower limit in order to obtain the numerical value on the left side of the equation. For an optically thin envelope
the condition $\tau < 1$ must be satisfied, imposing an upper bound to the black hole mass, namely
\begin{equation}
\frac{M}{M_\odot} < \frac{1.3\times 10^5}{n_{\infty}}
\end{equation}
where $n_{\infty}$ is the gas particle density far from the influence radius of the black hole. Hence, if accreting spherically, SMBHs at high-z with 
masses around $10^8-10^9 ~ M_\odot$ will necessarily have an optically thick envelope and a different inflow regime. An optically thick envelope
reduces the distance to the critical point and reduces also the accretion rate with respect to the optically thin case. In particular, when the radiation
field is quite important, a second critical point may exist in the flow besides the hydrodynamical one, according to Nobili et al. (1991). 

Accretion flows affected by radiation effects have been investigated by many authors in the past years (Maraschi et al. 1974; Flammang 1982; Milosavljevic et al.
2009). The radiation from the accreting envelope is essentially due to the free-free emission. For the optically thin case, the resulting luminosity 
is proportional to the square of the accretion rate and inversely proportional to the BH mass, i.e., $L \sim {\dot M^2}/M$. The flow becomes nearly
self-regulated when the optical depth of the infalling matter is greater than unity and under these conditions,
the luminosity approaches the Eddington limit (Milosavljevic et al. 2009). However, some authors claim that super-Eddington luminosities are possible
if the black-hole is embedded in a very dense gas that decreases the importance of radiation pressure effects (Pacucci et al. 2015). Super-Eddington
accretion rates were also found in some radiation-hydrodynamics simulations but based on one-dimensional geometry and particular conditions
of the ambient gas (Inayoshi et al. 2016). However, it is not certain whether such extreme conditions are sustainable considering
the violent environments of the first galaxies where the medium is affected by the star formation activity and supernovae.

If the BH is moving with respect to the gas, the situation is rather different. After passing the BH, a
conically shaped shock is produced in the flow in which the gas loses the momentum component perpendicular
to the shock front. After compression in the shock, gas particles within a certain impact parameter will fall into
the BH. One determinant factor describing the subsequent motion of the gas is the angular momentum. If the infalling gas has an specific angular momentum
$J$ that exceeds $2r_gc$, where $r_g = 2GM/c^2$ is the gravitational radius, the centrifugal forces will become important before the gas reaches 
the horizon. In this case, the gas will be thrown into near circular orbits and only after viscous stresses have transported away
the excess of angular momentum will the gas cross the BH horizon (Shvartsman 1971). In fact, the formation or not of a disk requires two conditions: the disk 
radius must be larger than the last stable circular orbit (equivalent to the condition $J > 2r_gc$) and must be
smaller than the typical dimension of the shock cone, e.g., $l_s \approx 2r_g(c/u_{\infty})^2$, where $u_{\infty}$ is the BH velocity with
respect to the gas. If the gas is highly turbulent, the velocity of eddies having a scale $k$ is given roughly by
\begin{equation}
V_t \sim V_0\left(\frac{k}{k_0}\right)^q
\end{equation}
In the case of a Kolmogorov spectrum, $q = 1/3$. However, it is more probable that the turbulent energy be dissipated mainly through shock waves and, in this case,
the spectrum is steeper with $q \sim 1$ (Kaplan 1954), a situation which will be assumed here. For our rough estimates, we adopt typical values for the turbulence
observed in our Galaxy, e.g., $V_0 \approx 10~ km.s^{-1}$, $k_0 \approx 10~ pc$ (Kaplan \& Pikel’ner 1970). The specific angular momentum associated with eddies 
is $J \sim V_tk$ and the specific angular momentum of the accreted gas corresponds to eddies of the order of twice the scale of the capture
impact parameter. Thus, the first condition for disk formation requires
\begin{equation}
M > 720\left(\frac{u_{\infty}}{50~km.s^{-1}}\right)^4 ~M_\odot
\end{equation}
whereas the second requires
\begin{equation}
M < 3.5\times 10^6\left(\frac{u_{\infty}}{50~km.s^{-1}}\right)^3~ M_\odot
\end{equation}
The conclusion of this brief analysis is that a disk can be formed after the shock front only if the BH mass is
in the range $10^3$ up to $10^6$ $M_\odot$. Notice that these values depend strongly on the black hole velocity.
Taking into account the restricted range of BH masses that allows the presence of a disk, the necessity of an adequate balance 
between the mass flow across the shock front and the flow throughout the disk that is controlled by viscous forces, as well as the variety 
of instabilities present in the flow after the shock. Under these conditions, the formation of a disk under these circumstances is rather uncertain. In 
this case, if a disk is not formed inside the accretion cone, the radiated luminosity is only a small fraction of the accretion power.

\section{Intermittent Growth of Black Holes}

As we have seen previously, the correlations between the black hole mass and the photometric or dynamical properties of the host galaxy suggest that the former
grows as the latter is assembled. The different physical processes involved on the growth of black holes inside galaxies require
a numerical treatment or, in other words, an appeal to cosmological simulations. In fact, there are different reasons justifying such an approach: first, because
a significant volume of the universe can be probed; second, because the dynamics of dark matter and the hydrodynamics of the gas, including physical processes like 
heating, cooling, ionization of different elements can be taken into account self consistently. Moreover, it is possible to study
environmental effects on the galaxies themselves as well as the chemical evolution of the interstellar and of the intergalactic gas, including the
effects of supernovae and turbulent diffusion of heavy metals. The simulations permit to test star formation recipes, models for the growth of seeds
and to investigate the influence of black holes on the environment when they are in an active phase.

\subsection{Cosmological Simulations}

The results described in this section were all derived from simulations performed at the Observatoire de la C\^ote d'Azur (Nice) in these past 
years. Details of the code and some results can be found, for instance, in the papers by Filloux et al. (2010, 2011) or Durier \& de Freitas Pacheco (2011).
For the sake of completeness, a short summary of the main features of the code will be presented here.

All the simulations were performed in the context of the $\Lambda$CDM cosmology, using the parallel TreePM-SPH code GADGET-2 in a formulation,
despite the use of fully adaptive smoothed particle hydrodynamics (SPH), that conserves energy and entropy
(Springel 2005). Initial conditions are established according to the algorithm COSMICS (Bertschinger 1995) and the evolution of the structures are
followed in the redshift range $60 \geq z \geq 0$.
Ionization equilibrium taking into account collisional and radiative processes were included 
following Katz, Weinberg \& Hernquist (1996), as well as the contribution
of the ionizing radiation background. The contribution of cooling processes such as collisional 
excitation of HI, HeI and HeII levels, radiative recombination, free-free
emission and inverse Compton were also included, using the results by Sutherland \& Dopita (1993). An 
interpolation procedure was adopted to take into account
the enhancement of the cooling as the medium is enriched by metals. The cooling functions computed by those authors 
are adequate for highly ionized gases
and for $T \geq 10^4 K$. At high redshifts ($100 > z > 20$) and inside neutral gas clouds a residual electron fraction
of about $n_e/n_H \approx 0.005$ is present (Peebles 1993) which is enough to act as a catalyst in chemical 
reactions producing molecular hydrogen. $H_2$ cooling
due to excitation of molecular rotational levels was introduced by using the results of Galli \& Palla (1998). After the appearance of the first stars, the 
gas is enriched by trace elements like O, C, Si, Fe, responsible for a supplementary cooling mechanism. The UV 
background with $h\nu < 13.6~eV$ is unable to ionize hydrogen
(and oxygen) in neutral gas clouds but it can ionize carbon, silicon and iron, which are mostly singly ionized
under these conditions. These ions have fine structure levels that can be excited by collisions either with 
electrons or atomic hydrogen, constituting an important
cooling mechanism at low temperatures, which was included in the code. The UV radiation from young massive stars able to 
ionize the nearby gas was computed for different
ionization species of hydrogen and helium, representing not only an additional (local) source of ionization but also of heating.

Feedback processes like the return of mass to the interstellar medium, supernova heating and chemical enrichment were all taken into account. The 
return of mass to the interstellar
medium was computed by assuming that the initial mass function (IMF) of stars with metallicities
$[Z/H] < −2.0$ is of the form $\zeta(m) \propto m^{-2}$ while stars more metal-rich are formed with a Salpeter IMF, e.g.,
$\zeta(m) \propto m^{-2.35}$. Stars in the mass range $40-80~ M_\odot$ leave a $10~ M_\odot$ black hole as a remnant, whereas a $1.4~ M_\odot$
neutron star is left if progenitors are in the mass range $9-40 M_\odot$ or a white dwarf remnant otherwise. The mass lost by the ''stellar particle" is 
redistributed according to the SPH kernel among the gas particle neighbors and velocities are adjusted in order to conserve the total
momentum in the cell. Moreover, the removed gas (except that ejected by supernovae, as discussed below) keeps its original chemical composition, contributing to
the chemical budget of the medium. In fact, AGB stars, planetary nebulae and WR stars enrich the medium in
He, C, N, but these contributions are not taken into account in the present version of the code.

Supernova explosions are supposed to inject both thermal and mechanical energy in the interstellar medium. Past investigations have shown that, when
heated, the nearby gas cools quite rapidly and the injected thermal energy is simply radiated away. However, when energy is injected in the form of kinetic
energy, the star formation process is affected (Navarro \& White 1993). In the present simulations, supernova explosions were supposed to inject 
essentially mechanical energy into the interstellar medium through a ''piston'' mechanism, represented by the momentum carried
by the ejected stellar envelope. The distance $D_p$ covered by such a “piston”, ejected with a typical velocity
$V_{ej} \sim 3000 km.s^{-1}$ in a time interval $\Delta t$ is $V_{ej}\Delta t$. Under this assumption a ''stellar cell'' is defined including
all gas particles inside a spherical volume of radius $D_p$ that will be affected by the ''piston''. The released
energy is redistributed non-uniformly among these particles. In this process, it is expected that the closest
gas particles receive more energy than the farthest ones. This was achieved by assigning to each gas particle $j$ a distance dependent 
weight $w_j(r) = A/r^n_{ij}$ , where $r_{ij}$ is the distance between the ''i-stellar" particle (host of the SN explosion) to the $j-gas$ particle 
inside the cell. The normalization constant is defined by $A = 1/\sum_j r^n_{ij}$ . Different values of the exponent (n = 2, 4) were tested.
Supernovae do not only contribute to the energy budget of the interstellar medium but also inject heavy
metals, leading to a progressive chemical enrichment of galaxies as well as of their nearby environment. Such a progressive enrichment was
treated by an adequate algorithm able to simulate the turbulent diffusion of metals trough the medium.

In the code, BHs are represented by collisionless particles that can grow in mass, according to specific rules that mimic accretion or merging 
with other BHs. Possible recoils due to a merging event and to the consequent emission of gravitational waves were neglected. BHs are 
assumed to merge if they come within a distance comparable to or less than the mean inter-particle separation. Seeds are supposed to have been 
formed from the first (very massive) stars and are supposed to have a mass of about 100 $M_\odot$. An auxiliary algorithm finds potential
minima where seeds are inserted in the redshift interval $15-20$.  A fraction of the energy released during the
accretion process is re-injected into the medium along two opposite “jets” aligned along the rotation axis of the disk, modeled by cones with an 
aperture angle of $20^o$ and extending up to distances of about 300 kpc. The adopted expression for the power of the jets is essentially that
given by the simulations of Koide et al. (2002).

\begin{figure}
\begin{center}
\rotatebox{0}{\includegraphics[height=9cm,width=12cm]{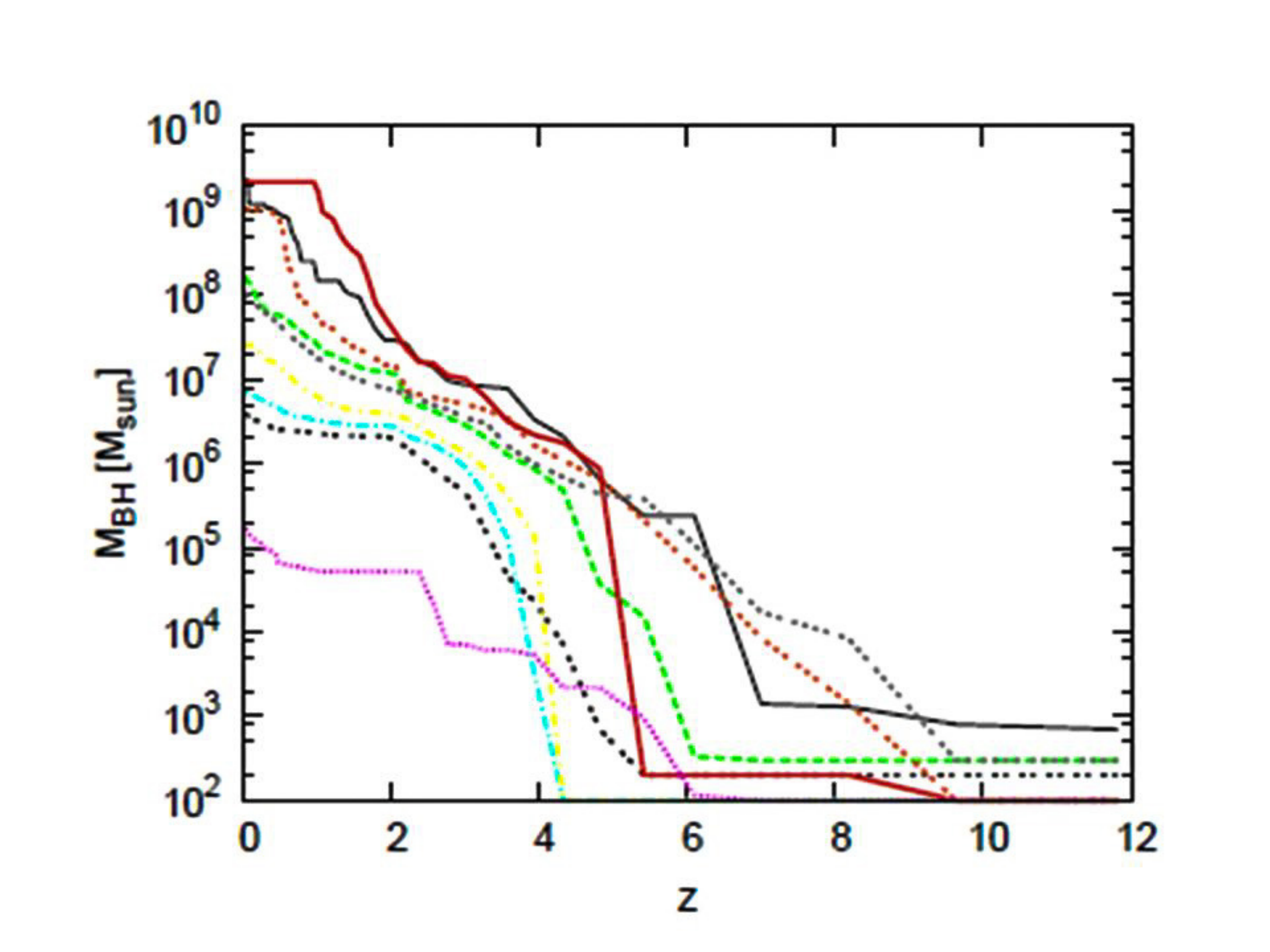}}
\end{center}
%\vfill
%\vspace{0.5cm}
%\vfill
\caption{Mass evolution of some individual black holes derived from simulations. Phases of activity or phases in which the black
hole is in a dormant state can be clearly identified.}
\end{figure}

\subsection{Properties of simulated SMBHs}

One of the main aspects of the growth of seeds by gas accretion concerns the fact that  masses do not increase continuously. In the hierarchical
model, galaxies are assembled in the filaments of the cosmic web or in the junction of filaments where clusters are formed. In filaments galaxies may 
capture fresh gas that will feed their central black holes. Gas may also come from merging events. However, from time to time the gas
in the central region is exhausted and the
process of growth stops until a new episode of gas capture is able to replenish the vicinity of the black hole. In fact, the amount of gas in the central
regions of the host galaxy is controlled by the capture processes and internal processes like star formation and feedback from supernovae and the black hole
itself. In figure 1 the individual
growth of some simulated black holes is shown. It is possible to verify that there are periods during which the black hole mass remains constant (case of
a ''dormant" black hole) and periods of gas accretion during which the black hole mass increases. In such a phase, the galaxy has an active nucleus, being
associated to an AGN or to a QSO. Notice that only at $z \leq 4$ some black holes having masses greater than $10^7~M_\odot$ have appeared. 

During the activity phase, the associated accretion disk is quite luminous and the luminosity depends essentially on the accretion rate. For a given
redshift it is possible to compute the total luminosity due to all active black holes and, consequently, to estimate the comoving luminosity density.
Such a luminosity density can be compared with observational data, permitting to test the robustness of the simulations. Figure 2 compares
the luminosity density evolution derived from simulations with data by Hopkins et al. (2007). The agreement is quite satisfactory suggesting 
that the main physical
aspects of the growth process are reasonably taken into account in the simulations.

\begin{figure}
\begin{center}
\rotatebox{0}{\includegraphics[height=9cm,width=12cm]{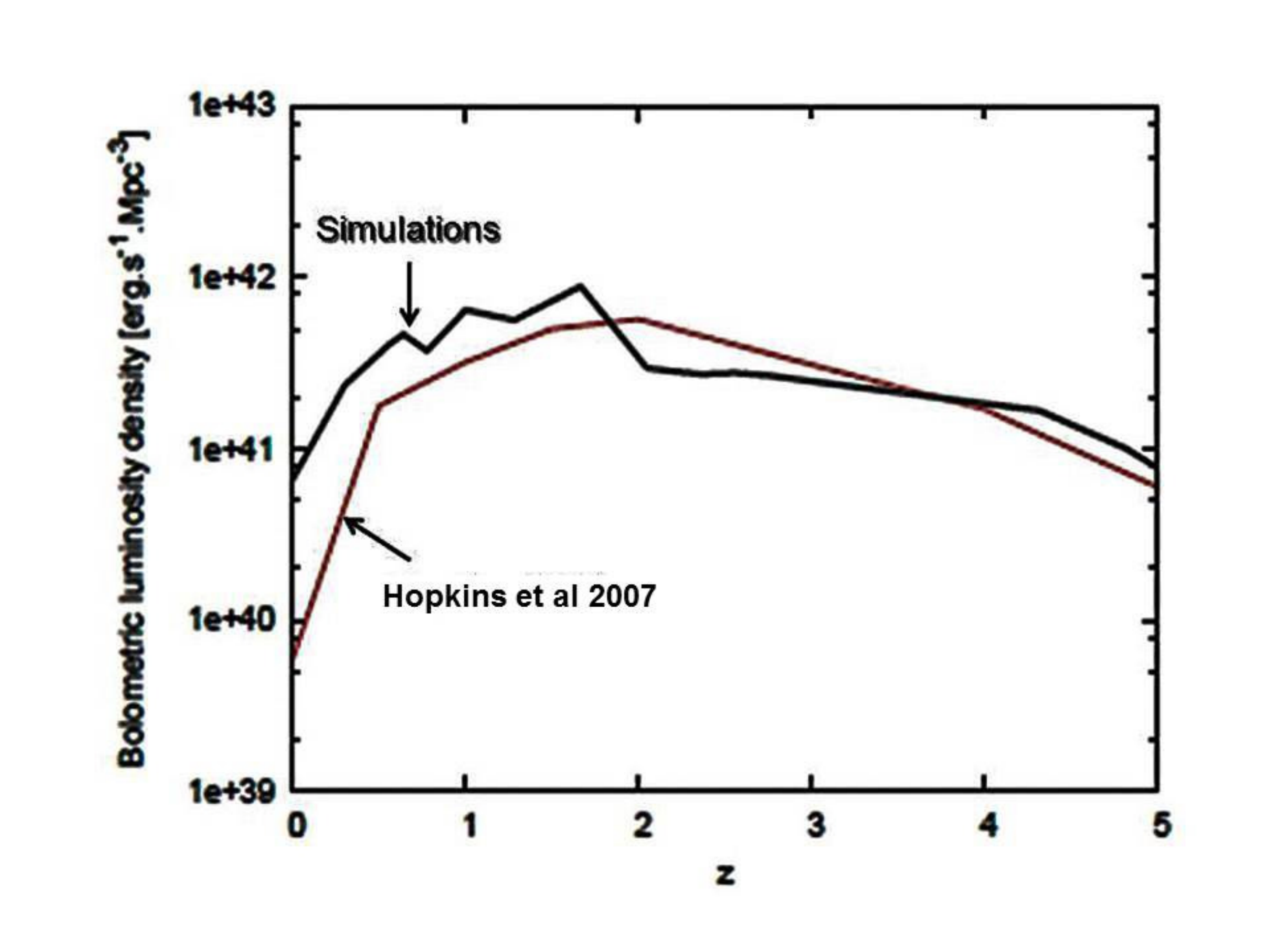}}
\end{center}
%\vfill
%\vspace{0.5cm}
%\vfill
\caption{Evolution of the luminosity density of active SMBHs derived from simulations compared with data on the luminosity 
density of QSOs}
\end{figure}

Another successful comparison concerns the relation between the present SMBH mass with the central velocity dispersion of galaxies projected
in the line of sight. This is done
in figure 3. Black squares represent the masses of SMBHs at z = 0 derived from simulations while red squares represent data taken from the literature.
There is a good agreement between simulated and observed data but some objects seem to have a higher black hole mass for the corresponding stellar
velocity dispersion of their host galaxies. In particular, this is the case for NGC 5252 and Cygnus A as it can be seen in the considered figure.
In our proposed scenario, these objects have not evolved intermittently but rather in a very short time scale in the early evolutionary phases of the 
universe, being presently the relics of such an active past.

If some SMBHs seem to have masses above that expected from the $M - \sigma$ relation as figure 3 suggests, there are other arguments indicating
that these objects followed indeed a different evolutionary path. At a given redshift, the simulations permit to compute the mass distribution
of SMBHs. This is shown in figure 4, which indicates that no SMBHs with masses above $10^7 ~ M_\odot$ is present at $z = 5$, in agreement with
the evolution of individual black holes shown in figure 1. This means that
the evolutionary path in which the BH mass grows as the host galaxy is assembled, which is probably the origin of the $M - \sigma$ relation, is unable to explain
the existence of very massive BHs in the early universe or SMBHs present today in bright galaxies like NGC 5252 or Cygnus A. In the next sections
an alternative evolutionary path will be examined. 
 
\begin{figure}
\begin{center}
\rotatebox{0}{\includegraphics[height=9cm,width=12cm]{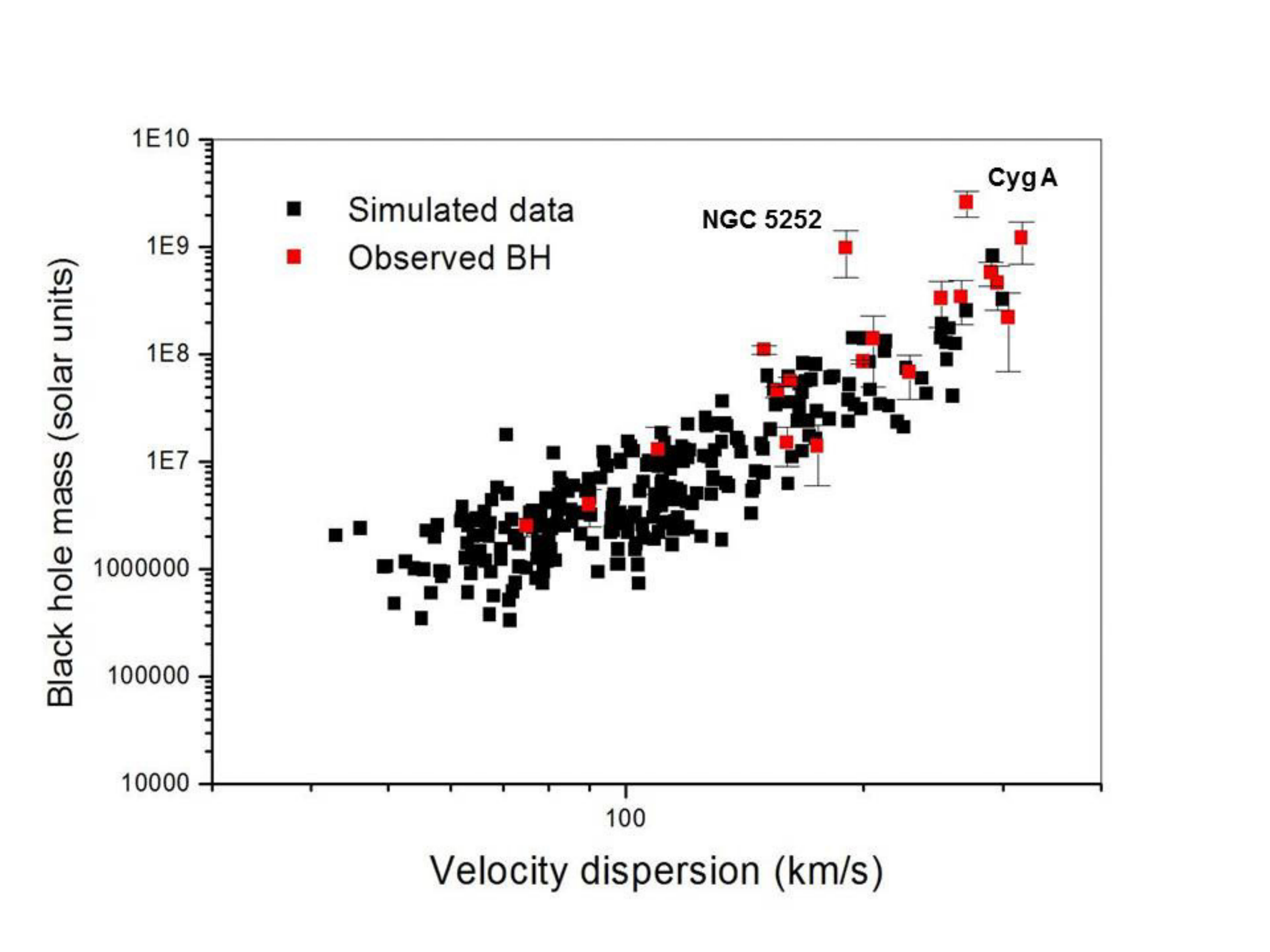}}
\end{center}
%\vfill
%\vspace{0.5cm}
%\vfill
\caption{Supermassive black hole mass versus projected central velocity dispersion of the host galaxy. Comparison between simulations
and data. Identified objects have probably evolved along a different path.}
\end{figure}

\section{The early formation of SMBHs}

As it was shown in the previous section, the coeval evolution of seeds and host galaxies is not able to explain the existence
of bright QSOs at high redshift and the fact that in the local universe some objects have masses higher than that expected
from the simulated $M - \sigma$ relation. 

Could a unique accretion disk form a SMBH in a timescale of about 1 Gyr? The answer to this question implies the solution of different
related problems. The disk must be quite massive in order to provide enough gas to form a $10^9 ~M_\odot$ black hole and the angular
momentum transfer must be very efficient in order to maintain a high accretion rate necessary to provide the observed luminosities
as well as a short timescale for the growth of the seed. In fact, numerical simulations suggest that after the merger of two galaxies, a 
considerable amount of gas is  settled into the central region of the resulting object. The gas loses angular momentum in a timescale 
comparable to the dynamical timescale (Mihos \& Hernquist 1996; Barnes 2002), forming circumnuclear self-gravitating disks having masses
in the range $10^6 - 10^9 ~ M_\odot$ and dimensions of about 100 - 500 pc. 

Massive accretion disks are, in general, self-gravitating in their early evolutionary phases, a situation that 
affects the usual dynamics of disks controlled only by 
the gravitational forces due to the central body. In fact, models of non steady self-gravitating accretion disks were computed
by Montesinos \& de Freitas Pacheco (2011, hereafter MP11) satisfying the aforementioned requirements or, in other words, they are luminous enough 
and they permit the growth of seeds in a short timescale, consistent with observations of bright QSOs at high $z$.
Some aspects of the work by those authors will be shortly reviewed below, followed by the presentation of new results.

\begin{figure}
\begin{center}
\rotatebox{0}{\includegraphics[height=9cm,width=12cm]{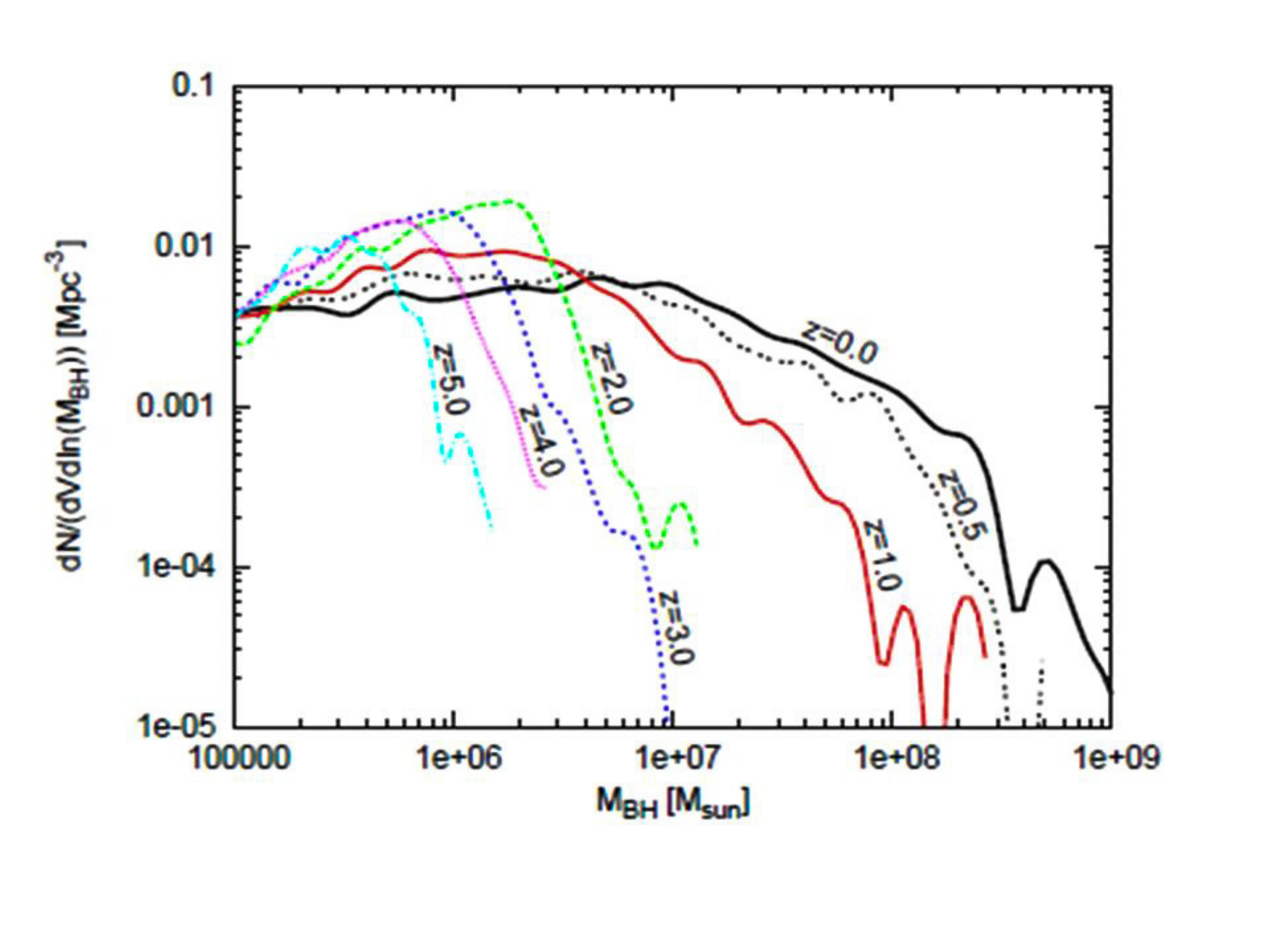}}
\end{center}
%\vfill
%\vspace{0.5cm}
%\vfill
\caption{Mass distribution of simulated black holes at different redshifts.}
\end{figure}

As mentioned before, the very early formation of a massive disk in the central region of a galaxy requires the presence of a large 
amount of gas. In fact, infrared sky surveys have discovered huge amounts of molecular gas (CO) at intermediate and high-z  QSOs 
(Downes et al 1999; Bertoldi et al 2003; Weiss et al 2007). In particular, the detection of CO emission in the quasar
J1148+5251 at $z = 6.42$ permitted an estimation of the molecular hydrogen mass present in the central region of the host galaxy
that amounts to $M(H_2) \sim 10^{10}~M_\odot$ (Walter et al. 2009). Moreover, at least in the case of the quasar J1319+0950 ($z = 6.13$)
there is a robust evidence that the gas is rotating (Shao et al. 2017), suggesting the presence of a gaseous disk. More 
recently, observations of J1342+0928 ($z = 7.54$)
indicate important amounts of gas and dust revealed by the infrared continuum and by the [CII] line emission (Venemans et al. 2017). All 
these observations support
the idea that some massive galaxies in their early evolutionary phases had large amounts of gas in their central regions, which could
have formed the massive accretion disks required by our model.

If observations seem to support the scenario in which massive disks fed the seeds of SMBHs, the other question concerns the
accretion timescale defining the growth of those seeds. The accretion rate is fixed by the mechanism of angular momentum transfer and depends
on the gas viscosity mechanism. Presently there is no adequate physical theory able to describe the gas viscosity in the presence of turbulent flows or in
the presence of magnetic fields.  The angular momentum transfer is generally described by the formalism introduced almost forty five years ago 
by Shakura \& Sunyaev (1973), in which the viscosity is due to subsonic turbulence and is parametrized by the relation $\eta = {\alpha}Hc_s$, where
$\eta$ is the viscosity, $\alpha \leq 1$ is a free parameter of the theory, $c_s$ is the sound velocity and $H$ is the vertical scale of height of the
disk. $H$ is supposed to be of the same order as the typical (isotropic) turbulence scale. However, disks based on such a formalism are, in general,
thermally unstable as demonstrated long time ago by Piran (1978). 

It is well known that self-gravitating disks may be also unstable but, in some cases, such an instability can be a source of turbulence in the flow 
(Duschl \& Britsch 2006). Simulations of the gas inflow in the central regions
of galaxies induced by the gravitational potential either of the stellar nucleus or the SMBH, reveal the appearance of
highly supersonic turbulence with velocities of the order of the virial value (Regan \& Haehenelt 2009; Levine et al. 2008; Wise, Turk \& Abel 2008). 
No fragmentation is observed in such a gas despite of being isothermal and gravitationally unstable. This behavior can be explained if
an efficient angular momentum transfer suppresses fragmentation. On the contrary, if the angular momentum transfer is inefficient, the 
turbulence decays and triggers global instabilities which regenerates a turbulent flow. Thus, one could expect that the flow will be self-regulated
by such a mechanism. In this case, the flow must be characterized by a critical Reynolds number ${\cal R}$, determined by the viscosity
below which the flow becomes unstable (de Freitas Pacheco \& Steiner 1973). This critical viscosity is given by
\begin{equation}
\label{visco}
\eta = \frac{2\pi r V_{\phi}}{{\cal R}}
\end{equation}
where $r$ is the radial coordinate and $V_{\phi}=r\Omega$ is the azimuthal velocity of the flow.

Another difference between ''$\alpha$"-disks and ''critical viscosity" models is the local balance of energy that is fixed by the equilibration of the rate 
of the dissipated turbulent energy with the radiated and advected energy rates or, in other words
\begin{equation}
\label{balance}
T_{r\phi}\frac{d\Omega}{dlg r} = \nabla\cdot F_{rad} + \varepsilon_{adv}
\end{equation}
In the above equation the left side represents the rate of turbulent energy dissipated per unit volume, the first term on the right side gives
the rate per unit volume of radiated energy and finally, the last term represents the rate per unit volume of advected energy. In eq.~\ref{balance},
$T_{r\phi}$ is the $(r,\phi)$ component of the stress tensor, $\Omega$ is the angular flow velocity, $F_{rad}$ is the radiative
flux and $\varepsilon_{adv}$ is the rate per unit volume of advected energy. In the ''$\alpha$"-disk model the considered stress component is given by
\begin{equation}
T_{r\phi}=\alpha\rho Hc_s\left(\frac{d\Omega}{dlg r}\right)
\end{equation}
while in the ''critical viscosity" model the stress is given by
\begin{equation}
T_{r\phi}=\frac{2\pi}{{\cal R}}\rho r^2\Omega\left(\frac{d\Omega}{dlg r}\right)
\end{equation}
The difference between the heating rates in both models implies that the temperature distribution along the disk is not the same and that the expected radiated
spectrum of each disk model is also different as it will be discussed later. 

In non-steady self-gravitating disk models, the dynamics of the disk evolves because the mass distribution changes with time as well as the mass 
of the central black hole. Near the BH the velocity is approximately Keplerian while beyond the transition region, where self-gravitation dominates, 
the rotational velocity decreases with distance more slowly than $1/\sqrt{r}$. Along the vertical axis, the disk is supposed to be in
hydrostatic equilibrium. The vertical scale of height varies as the disk evolves. At early phases the disk is geometrically thick in the central regions
due to radiation pressure effects. At late phases, the vertical scale of height increases with distance, approaching the behavior displayed by
canonical non self-gravitating models. Additional details, including a description of the numerical code used to solve the hydrodynamic equations can be found 
in MF11 as mentioned previously. 

\begin{figure}
\begin{center}
\rotatebox{0}{\includegraphics[height=9cm,width=12cm]{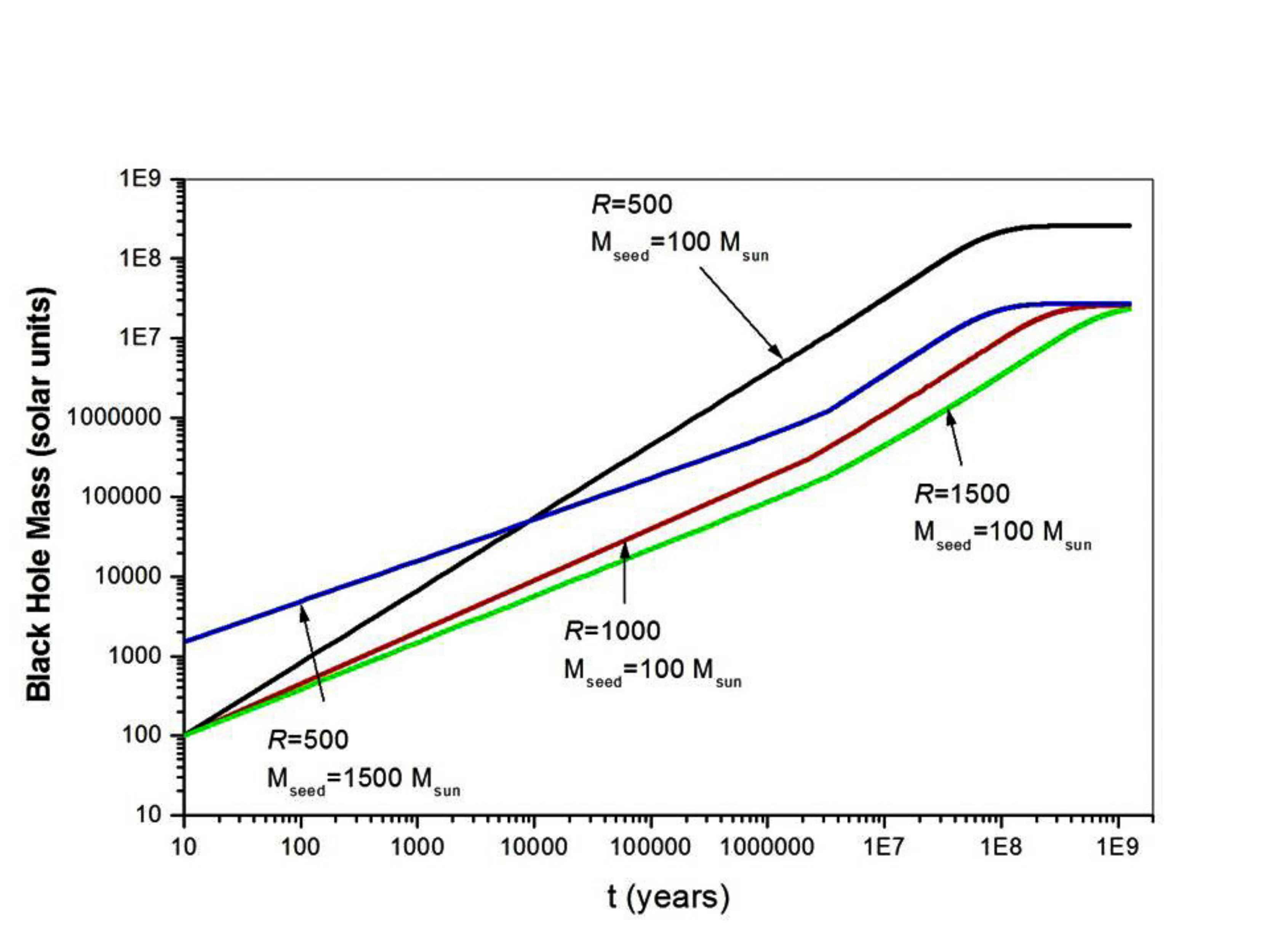}}
\end{center}
%\vfill
%\vspace{0.5cm}
%\vfill
\caption{Mass evolution of supermassive black holes for different values of the seed and the critical Reynolds number.}
\end{figure}

Figure 5, adapted from MF11, shows some examples of models characterized by different values of the seed (100 or 1500 $M_\odot$) and
of the critical Reynolds number (500, 1000 or 1500). Notice that an initial black hole of 100 $M_\odot$ can grow up to 
$3 \times 10^8 ~M_\odot$ in a timescale of only $10^8$ yr if the critical Reynolds number is 500 (black curve). Inspection of figure 5
shows that for the same
initial mass, if the Reynolds number is increased (red and green curves), the rate of growth decreases. This can be explained by the
fact that the accretion rate is inversely proportional to the viscous timescale, namely, $t_{vis}^{-1} \approx \eta/r^2 \approx \Omega/{\cal R}$.
This is an immediate consequence of the fact that increasing the Reynolds number it is more difficult to generate the turbulence. Therefore
the angular momentum transfer is less efficient, reducing the accretion rate. 

It is important to emphasize that while the in the internal parts of the disk the gas flows inwards, the outer parts expand 
as a consequence of the transfer of angular
momentum from inside to outside. Hence, only about 50\% of the initial mass of the disk is in fact accreted by the black hole. In the region where the
sign of the radial velocity changes (the ''stagnation" point) a torus-like structure is formed, supporting the scenario of the so-called
''unified model" of AGNs. The models by MF11 indicate that a substantial fraction of the expanding gas remains
neutral with a temperature in the range 100 - 2000 K most of the time. In the case of our own galaxy, such a behavior could be related
to the molecular “ring” of 2 pc radius observed around Sgr $A^*$ (Gusten et al. 1987). The physical conditions prevailing in the outskirts of the disk
are favorable to star formation and could be an explanation for the presence of massive early-type stars located in
two rotating thin disks detected in the central region of the Milky Way (Genzel et al. 2003; Paumard et al. 2006). 

\subsection{Further tests of the model}

In the very beginning of the disk evolution, the accretion rate (and the luminosity) increases very rapidly and then remains more or less 
constant during most of the growth process. At the final phases, the accretion rate decays very fast
once half of the disk mass is captured by the central black hole. Such a behavior can be seen in figures 2 and 3 of the paper by MF11.
Depending on the initial disk mass and on the critical Reynolds number,
the activity phase corresponding to the luminosity maximum  lasts for about $2\times 10^7$ up to $3\times 10^8$ years. 

Despite the fact that the accretion rate (and the luminosity) varies very little during the active phase, the spectral distribution of the radiation
emitted by the disk evolves. Such a spectral evolution is due to time variations of the optical depth radial profile as well as to time variations of the 
radial temperature distribution, as mentioned earlier. There is a continuous shift of the emission maximum toward longer 
wavelengths that is a consequence of the decreasing average disk temperature as a function of time. 
In general for wavelengths $\lambda \geq 0.15~ \mu m$ the spectral intensity can be well represented by a power-law, that is, $I_{\lambda} \propto 
\lambda^{-\alpha}$, where the power index is in the range $0.9 < \alpha < 1.3$, in agreement with values derived from most of quasar spectra.

The modeling of the spectral emission of the disk permits an estimate of the bolometric correction. Usually, the luminosity
in a given wavelength is derived from observations of monochromatic fluxes and luminosity distances, which depend on the redshift. The bolometric luminosity 
can be computed by adopting an adequate correction. Nemmen \& Brotherton (2010) have estimated the bolometric correction for luminosities 
at $\lambda = 0.30 \mu m$ ~ based on models by Hubeny et al. (2000).
The grid of models by the latter authors assimilates to each annulus of the disk an effective temperature and gravity, which are used to compute the emergent
spectrum of an equivalent stellar atmosphere defined by those parameters. The sum of the radiation from all annuli gives the resulting spectrum of the 
disk. However the effective temperature formula adopted by those authors is adequate for a steady disk whose dynamics is dominated by the central black hole.
In the case of non-steady self-gravitating disks the situation is rather different because both the local gravity and the effective temperature vary with time.
Fortunately, the bolometric corrections for the monochromatic luminosities at $\lambda = 0.30 \mu m$ ~ or at $\lambda = 3.6 \mu m$~ do not vary too 
much during the active phase and a suitable average correction can be defined.

\begin{figure}
\begin{center}
\rotatebox{0}{\includegraphics[height=9cm,width=12cm]{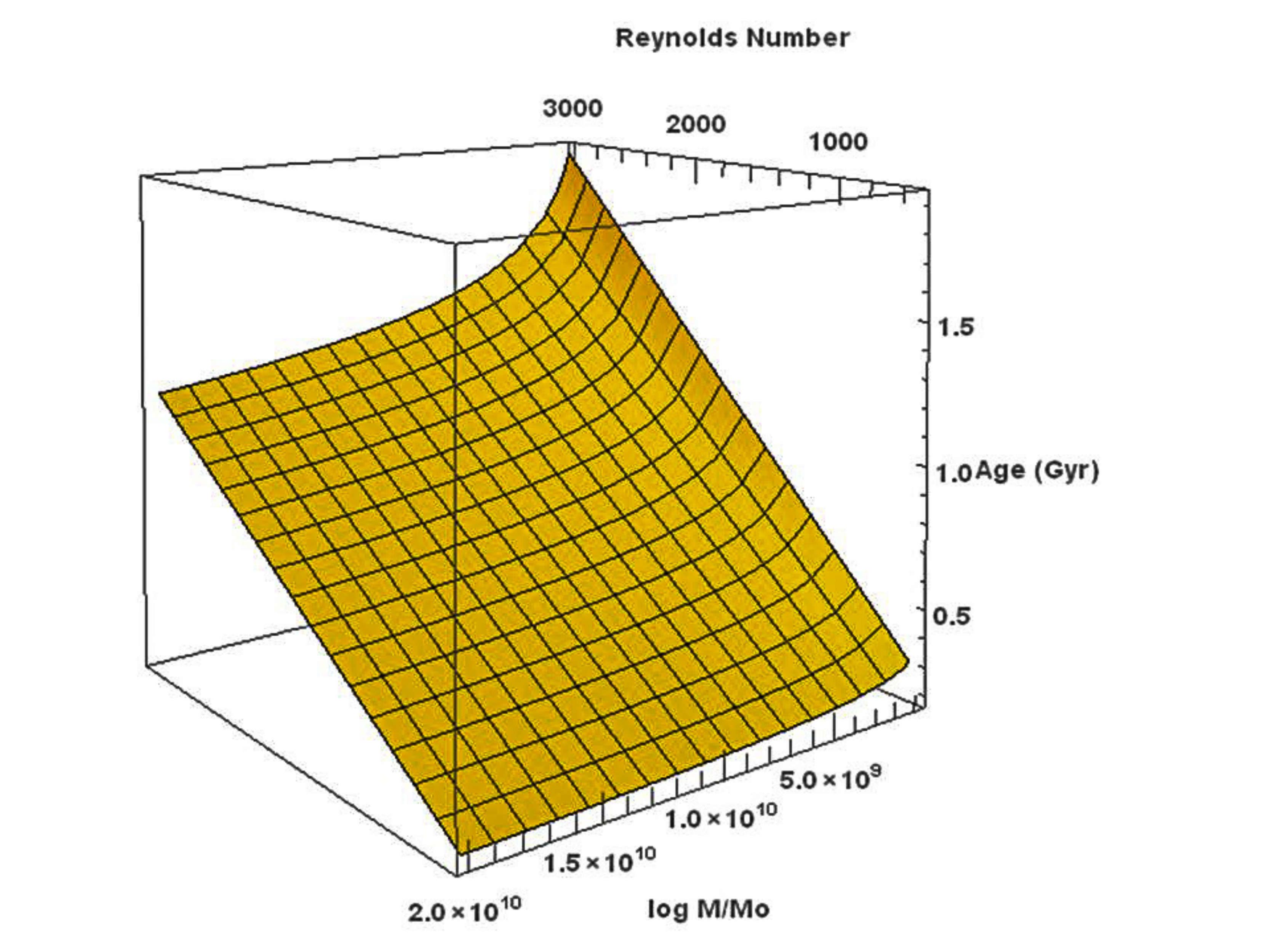}}
\end{center}
%\vfill
%\vspace{0.5cm}
%\vfill
\caption{Surface defining the space of parameters, mass of the black hole, age and critical Reynolds number.}
\end{figure}

The adopted procedure to estimate such a bolometric correction requires an adequate choice of the representative parameters of the disk, since the
seeds must be able to grow in timescales less than 1 Gyr and form SMBHs with masses larger than $5\times 10^8 ~ M_\odot$. 
Figure 6 shows the surface ''M-age-${\cal R}$" derived from a grid of models where the seed mass was fixed
to 100~$M_\odot$. It worth mentioning that in such a plot the parameter ''age" means the timescale required for the seed to accrete 50\% of the
initial disk mass, the same definition that was adopted by MF11. Inspection of figure 6 indicates that
Reynolds numbers in the range $1000 < {\cal R} < 2500$ are required in order to satisfy those constraints.  Then, bolometric corrections 
at $\lambda$ = 0.30~$\mu m$ and at $\lambda$= 3.6~$\mu m$  were computed for a series of models characterized by ${\cal R}$ = 2200, mass of the seed
equal to 100~$M_\odot$  and different initial 
disk masses, corresponding to about twice the final black hole masses. After averaging the results from different models, the corrections are simply given by

\begin{equation}
\log L_{bol} = \log \lambda L_{\lambda} + 0.83 \,\,\,\, for\,\, \lambda = 0.30~\mu
\end{equation}
and
\begin{equation}
\log L_{bol} = \log \lambda L_{\lambda} + 0.92 \,\,\,\, for\,\, \lambda = 3.6~\mu m
\end{equation}
where the luminosities are given in $erg.s^{-1}$. In figure 7 the bolometric correction for $\lambda$ = 0.30~$\mu m$ derived from these models is compared with the correction adopted by Nemmen \& Brotherton (2010) based on steady and non self-gravitating disk models. It should be emphasized that the present bolometric luminosities 
derived either from UV or infrared 
monochromatic luminosities are in very good agreement when the corrections above are applied.

The present disk model can also be tested by the comparison between theoretical predictions and data in the diagram bolometric luminosity versus 
black hole mass. Data on high redshift QSOs ($z \geq 6.0$), including masses, monochromatic luminosities at $\lambda$ = 0.30 or 3.6 $\mu m$~ and redshift, 
compiled by Trakhtenbrot et al. (2017) were used in the calculations. Black hole masses were estimated from the width of MgII lines and monochromatic UV-luminosities 
were derived from the best-fit model of the Mg II emission line complex. Then, using the derived bolometric corrections,
the bolometric luminosity of each object was estimated and plotted as a function of the black hole mass in figure 8. The ''Mass-Luminosity" relation derived from our models
can be adequately represented by the fit 
\begin{equation}
\label{luminosity}
\frac{L_{bol}}{L_\odot} = 1.41\left(\frac{500}{{\cal R}}\right)\left(\frac{M_{seed}}{100M_\odot}\right)^{0.52}\left(\frac{M}{M_\odot}\right)^{1.5}
\end{equation}
which depends essentially on the seed mass and on the critical Reynolds number. Such a relation for $M_{seed} = 100~M_\odot$ and ${\cal R}$ = 2200 is shown in
figure 8 as a red line. 
\begin{figure}
\begin{center}
\rotatebox{0}{\includegraphics[height=9cm,width=12cm]{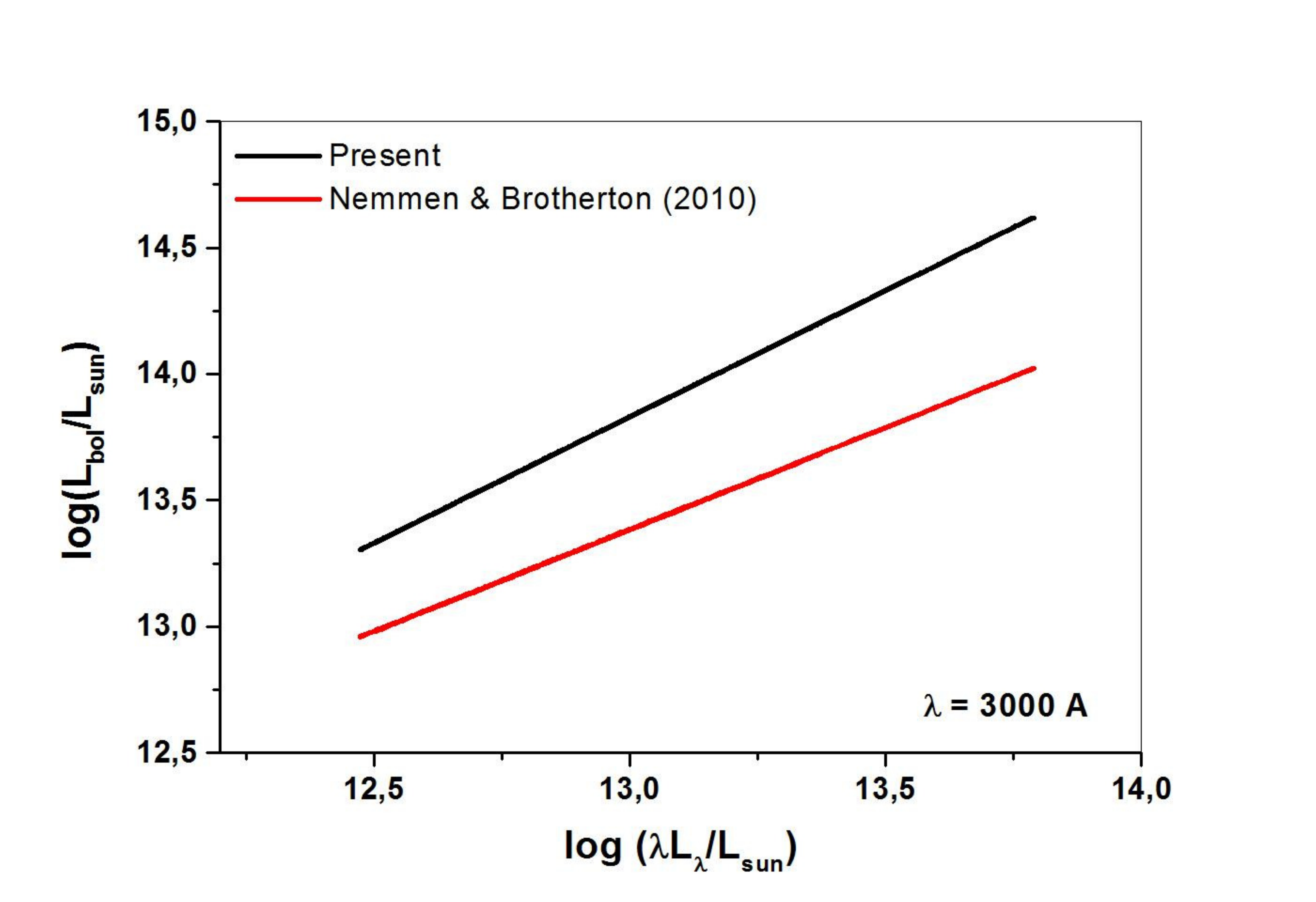}}
\end{center}
%\vfill
%\vspace{0.5cm}
%\vfill
\caption{Comparison between bolometric corrections based on steady and non-steady accretion disks.}
\end{figure}

In figure 8 is also shown the expected relation for the Eddington limited luminosity (see eq.~\ref{eddington}), frequently used to estimate the mass of
the black hole. Notice that the theoretical ''M-L" relation approaches the Eddington limit for black holes masses greater than $10^{10}~M_\odot$. 
It is worth mentioning that identifying 
the bolometric luminosity with the Eddington limit leads to an underestimate of the black hole mass, as it can be seen in the considered plot,
where the majority 
of the data points are below the expected Eddington limit line. On the other hand,
the theoretical ''L-M" relation displayed in figure 8 shows clearly that our disk models radiate
below the Eddington limit. 

It should be emphasized that in the case of accretion 
disks, the balance between gravity and radiation pressure along the vertical axis must be considered locally. The disk is locally stable if the radiative 
flux along the vertical direction is not greater than a critical flux limit given by 
\begin{equation}
\label{critical}
F_{rad} = \frac{\sqrt{3}}{4\pi}\left(\frac{m_Hc}{\sigma_T}\right)\left(\frac{\tau_s}{\tau_{ff}}\right)^{1/2}g_z
\end{equation}
In the above equation $\tau_s$ is the optical depth due to electron scattering, $\tau_{ff}$ is the optical depth due to free-free absorption and
$g_z$ is the local vertical gravitational acceleration. The condition 
expressed by eq.~\ref{critical} is valid in the inner regions of the disk where the electron scattering dominates over free-free absorption and
where radiation pressure effects are more important. When the vertical radiation flux is higher than the critical value, the hydrostatic equilibrium
is destroyed and outflows can be generated. Three-dimensional radiation magneto-hydrodynamical simulations were performed by Jiang et al. (2017), who
have studied the evolution of an accretion disk with torus centered on a $5\times 10^8~M_\odot$ black hole. The radiation pressure in the internal
regions of the disk may reach values up to $10^6$ times the gas pressure under certain conditions, producing outflows. In these simulations, the
angular momentum transfer is controlled by magnetohydrodynamic turbulence that is not the case of our models. 

\begin{figure}
\begin{center}
\rotatebox{0}{\includegraphics[height=9cm,width=12cm]{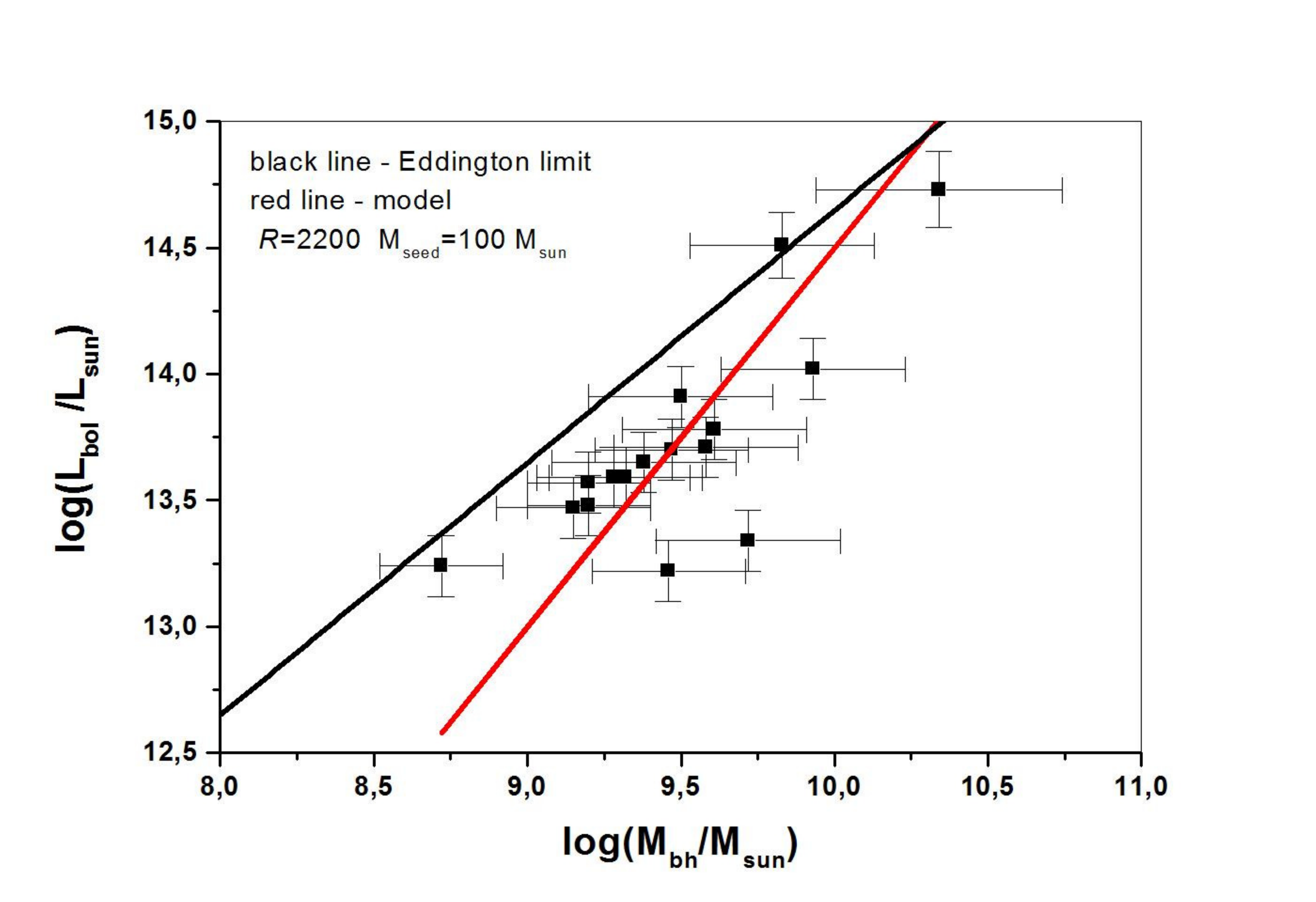}}
\end{center}
%\vfill
%\vspace{0.5cm}
%\vfill
\caption{Bolometric luminosity plotted versus the black hole mass for objects with $z \geq 6.0$. The bolometric correction at 0.30~$\mu m$~is
given in the text. The red line corresponds to eq.~\ref{luminosity} with ${\cal R}$=2200 and $M_{seed} = 100~M_\odot$. The black line corresponds
to the Eddington limit.}
\end{figure}

The present accretion disk models can be also tested in the diagram black hole mass versus age (figure 9). This plot is simply the projection of the
surface displayed in figure 6 on the considered plane ''M-age". Since the age parameter defined above is not directly accessible from observations, the age of 
the universe derived from the observed redshift of the QSO was used to plot the objects listed by Trakhtenbrot et al. (2017). The age of the universe
represents a robust upper limit to the age parameter of the model.
Theoretical predictions shown in figure 9 (solid lines) were computed for the same value of the seed mass ($M_{seed}$ = 150~$M_\odot$) and for 
two different critical Reynolds number: ${\cal R}$ = 1800 and ${\cal R}$ = 2500. These two values enclose in the plot most of the observed high-z QSOs,
strongly constraining this fundamental parameter of the model.

\begin{figure}
\begin{center}
\rotatebox{0}{\includegraphics[height=9cm,width=12cm]{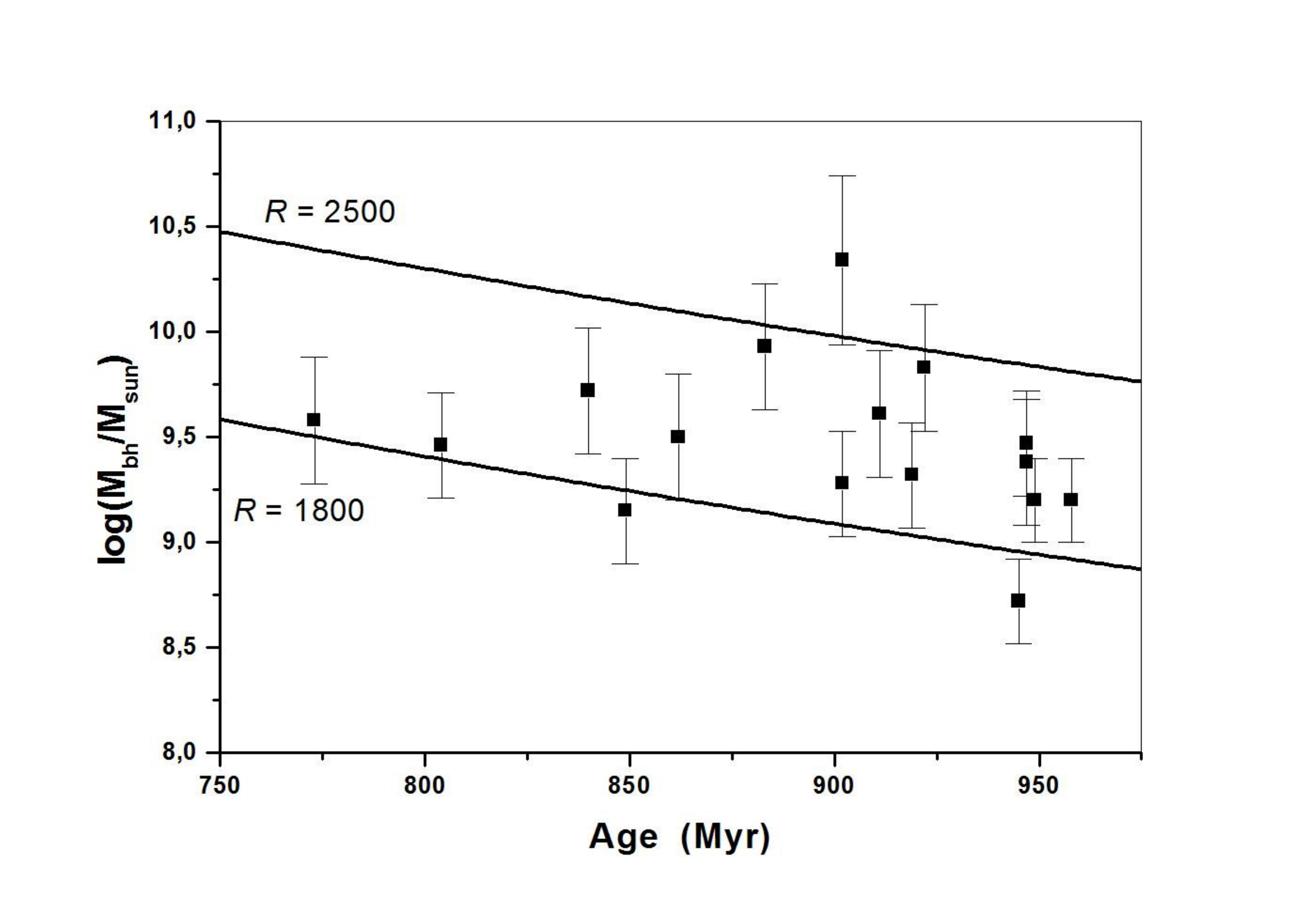}}
\end{center}
%\vfill
%\vspace{0.5cm}
%\vfill
\caption{Black hole masses versus age of the universe. Solid lines represent theoretical predictions from non-steady
self-gravitating accretion disks.}
\end{figure}

\section{Conclusions}

Present astronomical data are not in contradiction with a scenario in which two different evolutionary paths exist for the
formation of SMBHs from small mass seeds. 

In the first evolutionary path, seeds having masses around 100~$M_\odot$ grow intermittently following the gradual assembly of the host galaxy,
according to the hierarchical picture. In this case, the coeval evolution between the host galaxy and the seed must be investigated
by cosmological simulations. This procedure is justified by the complexity of the physical mechanisms involved in the process of  growth.
As we have previously seen, these numerical experiments are able to reproduce the observed luminosity density of QSOs and the
observed correlations between the black hole mass at $z =0$ and the properties of the host galaxy like the stellar luminosity or the central 
projected stellar velocity dispersion. Despite these successful results,
these simulations are unable to form SMBHs with masses around $10^9~M_\odot$ at high
redshift, unless the masses of the seeds are dramatically increased up $10^5 - 10^6 ~ M_\odot$. Although this could be a possibility and despite
some studies in this sense, such an alternative seems to be unrealistic.  

The existence of bright QSOs at $z \approx 6 - 7$ and the difficulty for cosmological simulations to form these objects points toward a new
direction, that is, the possibility of a very fast growth of seeds fed by massive accretion disks. This picture is
supported by observations of large amounts of gas and dust in QSOs at high-z as discussed before. Models of non-steady self-gravitating disks
in which the angular momentum transfer is controlled by turbulent viscosity were developed by Montesinos \& de Freitas Pacheco (2011). These
models have demonstrated that seeds can grow in timescales of the order of 1 Gyr or even less, being able to explain the main features of QSOs
observed at high-z.

Further investigations on these ''critical-viscosity" disks permitted an estimation of the bolomentric correction that should be applied to
monochromatic luminosities measured at 0.30~$\mu m$ and 3.6~$\mu m$. These corrections permitted a comparison of existing data with theoretical
predictions in the diagram ''Luminosity-Mass". Such a plot strongly suggests that accretion disks radiate below the so-called Eddington limit.
This means that black hole masses derived from such a limit are underestimated. Another useful diagram permitting the comparison
of model predictions with data is the plot ''Mass - age". Here it is necessary to recall the remarks done
before, that is, the age derived from the redshift is the age of the universe at that moment, representing only a robust upper limit to the disk age.
Nevertheless, despite such limitations, both diagrams permit to constrain the two important parameters of the model, namely, the mass of the seeds and the 
critical Reynolds number. The former is probably in the range 100-150~$M_\odot$ while the latter should be in the interval $1800 < {\cal R} < 2500$.

Finally, it worth mentioning that some SMBH in the local universe ($z = 0$) have masses above that expected from simulations. This is the case
of NGC 5252 and Cynus A as already mentioned, but may also be the case of NGC 3115 and probably of NGC 4594. This last object is more
uncertain since its estimated mass is only 4.4 times greater that that expected from the simulated ''M-$\sigma$" relation. These objects are
probably the remnants of a fast growth occurred in the early evolutionary phases of the universe and not the consequence of a coeval
evolution involving the seed and the host galaxy.

\end{document}